\documentclass{epl}

\title{Double stochastic resonance peaks in systems with dynamic phase transitions}
\shorttitle{Double SR peaks in systems with dynamic phase transitions}

\author{Beom Jun Kim\inst{1}\thanks{E-mail: \email{kim@tp.umu.se}} \and P. Minnhagen\inst{1} \and 
Hyun Jin Kim\inst{2} \and M. Y. Choi\inst{2} \and Gun Sang Jeon\inst{3}}
\shortauthor{B.J. Kim \etal}
\institute {
   \inst{1} Department of Theoretical Physics, Ume{\aa} University,
            901 87 Ume{\aa}, Sweden \\
   \inst{2} Department of Physics, Seoul National University, 
            Seoul 151-742, Korea \\ 
   \inst{3} Center for Strongly Correlated Materials Research,
            Seoul National University, Seoul 151-742, Korea
}
\pacs{05.40.$-$a}{Fluctuation phenomena, random processes, noise, and Brownian motion}
\pacs{74.50.$+$r}{Proximity effects, weak links, tunneling phenomena, and Josephson effects}
\pacs{75.10.Hk}{Classical spin models}

\begin{document}

\maketitle

\begin{abstract}
To probe the connection between the dynamic phase transition and
stochastic resonance, we study the mean-field
kinetic Ising model and the two-dimensional Josephson-junction array
in the presence of appropriate oscillating magnetic fields.
Observed in both systems are
{\it double} stochastic resonance peaks,
one below and the other above the dynamic transition temperature,
the appearance of which is argued to be a generic property
of the system with a continuous dynamic phase transition.
In particular, the frequency matching condition around the dynamic
phase transition between the external drive frequency and the
internal characteristic frequency of the system is identified
as the origin of such double peaks.
\end{abstract}

When a system with an energetic activation barrier is weakly coupled
to a temporally periodic driving, the inherent thermal 
stochastic noise can enhance the signal out of the system rather than
weaken it.  This behavior, called stochastic resonance (SR)~\cite{gammaitoni},
has been studied extensively in systems with a few degrees of freedom;
the appearance of a resonance peak at a finite temperature $T_{SR}$,
characteristic of SR,
is conveniently explained by matching the two time scales:
the deterministic one due to the external driving and 
the stochastic one which is inversely proportional to the 
thermal activation rate (Kramer rate)~\cite{gammaitoni}.
Recently, there have also been attempts to investigate the SR 
in more complex systems~\cite{lindner,neda,leung,bjkimJJL}.
Here one should note that such a system with many degrees of freedom 
may undergo a phase transition at a nonzero temperature 
and have the characteristic time scale
not described by the simple Kramer rate~\cite{nicolis}. 
Whereas the Kramer rate gives a diverging time scale only at zero temperature, 
the system with a finite-temperature continuous transition possesses 
a diverging characteristic time scale at a nonzero critical temperature $T_c$, 
yielding finite values of the characteristic time
{\em both} below {\em and} above $T_c$. 

In this Letter we propose to extend the time-scale matching condition
in such a way that the matching between the time scale of the
external periodic driving and the characteristic time scale associated with 
the phase transition determines the position of the SR peak. 
Since the time scales then match at two distinct temperatures, 
one below and the other above $T_c$,
it follows that the system with a continuous phase transition 
should show {\it double} SR peaks, 
one below and the other above the transition. 
In order to elucidate this scenario, we
investigate the interplay between the dynamic phase 
transition and the SR behavior in
the infinite-ranged kinetic Ising model and in the two-dimensional (2D) 
fully frustrated Josephson-junction array (FFJJA),
driven by appropriate oscillating magnetic fields.
Unlike the latter system, where the SR behavior has not been examined, 
the kinetic Ising model in an oscillating magnetic field 
has been studied previously via the mean-field approximation 
and numerical simulations,
with regard to both the dynamic phase transition~\cite{chak,tome90,acharyya}
and the SR behavior~\cite{neda,leung},
and double SR peaks have been found and discussed by Leung and N{\'e}da in 
\cite{leung}.
However, the physics behind the emergence of double peaks has not been 
addressed per se:
To our knowledge, 
neither the connection between the origin of the double peaks and
the existence of a dynamic phase transition nor the generality
of this connection has been recognized and addressed at all.

We begin with the infinite-ranged kinetic Ising model described by
the Hamiltonian
\begin{equation}
H = -\frac{J}{N}\sum_{i<j} \sigma_i \sigma_j - h(t)\sum_{i=1}^N \sigma_i, 
\end{equation}
where the Ising spin $\sigma_i$ at site $i$ can have values $\pm 1$,
$J$ is the coupling strength, 
and the external oscillating magnetic field is given by
$h(t) = h_0 \cos \Omega t $. 
Henceforth we set $J\equiv 1$ and measure the temperature $T$ in units of $J/k_B$.
For the above infinite-ranged Hamiltonian, the mean-field approximation is exact
and yields the equation of motion under the Glauber dynamics~\cite{leung,tome90}:
\begin{equation} \label{eq:glauber}
\frac{dm(t)}{dt} = -m(t) + \tanh\left( \frac{m(t) + h(t)}{T} \right), 
\end{equation}
where the magnetization $m(t)$ is given by the ensemble average of the spin 
at time $t$.

We first recapitulate the dynamic phase transition present in the system described by
Eq.~(\ref{eq:glauber}).
In the absence of the external field, the system undergoes 
a thermodynamic phase transition at the transition temperature $T_0=1$ 
between the low-temperature ferromagnetic phase $(m_0 \neq 0)$
and the high-temperature paramagnetic phase $(m_0 =0)$,
where $m_0$ is the spontaneous magnetization in equilibrium. 
In the presence of the external oscillating field, on the other hand, 
the system displays a dynamic 
phase transition~\cite{leung,chak,tome90,acharyya},
for which the dynamic order parameter $Q$ can be defined to be
\begin{equation} \label{eq:op}
Q \equiv \lim_{n \rightarrow \infty} Q_n 
\end{equation}
with $Q_n$ denoting the time average during the $n$th period
of external driving:
\begin{equation} \label{eq:Qn}
Q_n \equiv \frac{\Omega}{2\pi}\int_{2\pi n/\Omega}^{2\pi(n+1)/\Omega} dt\, m(t) .
\end{equation}
Note that $Q$ separates the ferromagnetic phase $(Q\neq 0)$ from
the paramagnetic phase $(Q=0)$ and
determines the dynamic phase transition temperature $T_c$.
The phase boundary, obtained numerically
by solving Eqs.~(\ref{eq:glauber})-(\ref{eq:Qn}) \cite{tome90}
and analytically via the adiabatic approximation valid for sufficiently small
$\Omega$~\cite{SRlong}, 
in general connects the points $(T, h_0)=(1, 0)$ and $(0, 1)$ on the $T$-$h_0$ plane. 
For given frequency $\Omega$, as the amplitude $h_0$ is increased from zero,
the transition not only occurs at a lower temperature, lowering $T_c$,
but also changes its nature from a continuous type to a discontinuous one.
The division point at which nature of the transition changes
moves along the phase boundary from $(1, 0)$ towards $(0, 1)$ 
as $\Omega$ is increased from zero. 
Good overall agreement exists in a broad range of temperatures 
between numerical results and analytical ones via the adiabatic approximation
and the perturbation expansion~\cite{SRlong}.

\begin{figure}
\onefigure{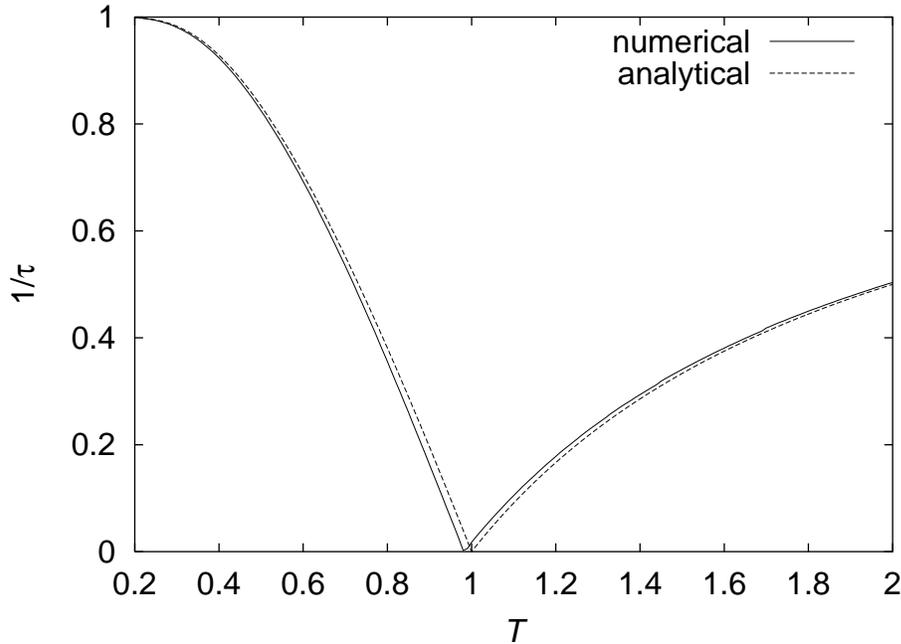}
\caption{
Inverse of the relaxation time versus temperature, 
computed numerically from the behavior of $Q_n$ 
for $h_0 = 0.1$ and $\Omega/2\pi = 0.1$ (full line)
and analytically via the linear response theory (dashed line).
The relaxation time is shown to diverge, 
as the transition temperature is approached either from
above or from below.
}
\label{fig:tau}
\end{figure}

A key quantity in our argument for the double SR 
peaks is the relaxation time $\tau$ or equivalently, the intrinsic frequency scale 
$1/\tau$ of the system, which describes the decay of the average magnetization: 
$Q_n -Q \propto e^{-t/\tau}$. 
Thus one may start from, e.g., the ordered state $m(t{=}0) = 1$, and extract
the exponential decay of $Q_n$ to obtain numerically the relaxation time~\cite{acharyya}.
At weak driving fields, 
the relaxation time can also be calculated analytically
to the linear order in the amplitude $h_0$~\cite{leung,SRlong}:
\begin{equation} \label{eq:tau}
\frac{1}{\tau} = 1 - \frac{1}{T}\frac{1}{{\rm cosh}^2(m_0/T)},
\end{equation}
where the spontaneous magnetization $m_0$ is determined by the equation of state:
$m_0$ $=$ $\tanh (m_0 /T)$. 
Figure~\ref{fig:tau} displays the relaxation time for $h_0 =0.1$ and
$\Omega/2\pi =0.1$, determined by these two methods. 
It is shown that at the dynamical phase transition the 
characteristic frequency $1/\tau$ vanishes or equivalently,
the relaxation time $\tau$ diverges. 
Note also the good agreement between the analytical and numerical results,
implying that Eq.~(\ref{eq:tau}) gives an excellent approximation, 
except for the discrepancy in the temperature at which $\tau$ diverges:
Equation (\ref{eq:tau}) leads to the divergence
at the equilibrium transition temperature $T_0=1$, whereas the 
actual relaxation time diverges at the dynamical transition temperature
$T_c\approx 0.98$, slightly lowered by the external driving field.
It is important here that a continuous {\em dynamical} transition
as well as a continuous equilibrium transition 
is associated with a divergent relaxation time.  
In particular the relaxation time can assume the same (finite) value 
at two temperatures, one above and the other below the transition.

\begin{figure}
\onefigure{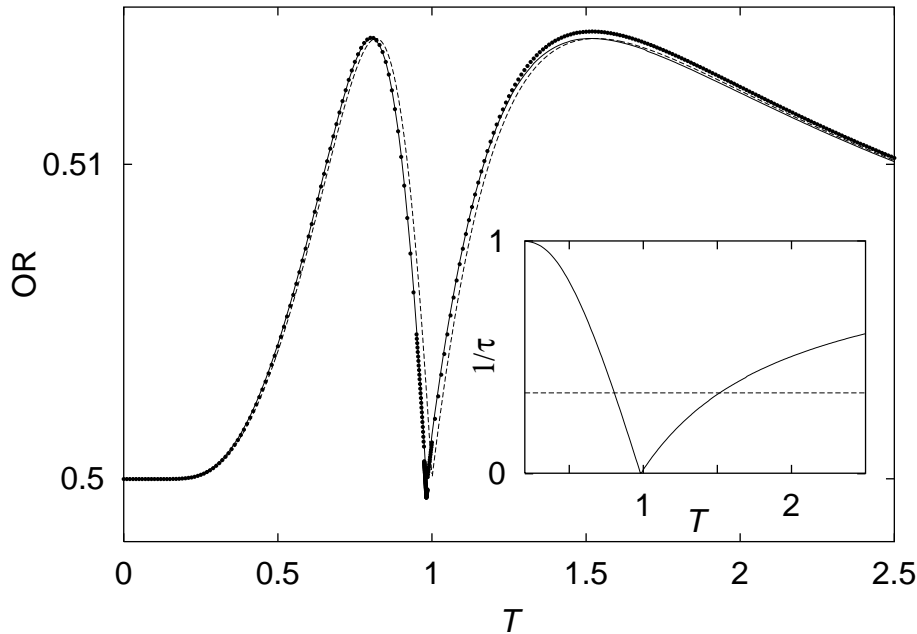}
\caption{
Occupancy ratio versus temperature in the kinetic Ising model 
for $h_0 = 0.1$ and $\Omega/2\pi = 0.1$.
Data directly from Eq.~(\ref{eq:OR}) together with
numerical integration of Eq.~(\ref{eq:glauber}) are represented by
small circles whereas the analytical expression given by Eq.~(\protect\ref{eq:ORanal})
with $\tau$ obtained numerically in Fig.~\ref{fig:tau} 
is plotted by the solid line.
The dashed line corresponds to the linear response calculation
[Eq.~(\protect\ref{eq:ORanal}) with $\tau$ given by Eq.~(\ref{eq:tau})].
Inset: the time scale matching condition with the solid line corresponding 
to the inverse relaxation time obtained numerically in Fig.~\ref{fig:tau}
and the horizontal dashed line 
to the right-hand side in Eq.~(\ref{eq:matching}).
The two crossing points below and above the transition temperature
determine the positions of the two SR peaks in the main body.
}
\label{fig:ORanal}
\end{figure}

It is convenient to study the SR behavior in the system by means of
the occupancy ratio (OR), defined to be the average fraction 
of the spins in the direction of the external field~\cite{bjkimJJL}:
\begin{equation} \label{eq:OR}
{\rm OR} \equiv \left\langle \frac{ \mbox{number of spins in the
direction of $h(t)$}} {\mbox{total number of spins} } \right\rangle.
\end{equation}
At sufficiently low temperatures most spins are stuck to one of
two spin directions ($\pm 1$) and  do not follow
the direction preferred by the field $h(t)$. 
Since $h(t)$ oscillates with period $2\pi/\Omega$,
most spins point in the same direction as $h(t)$ during one half of the period 
while almost no spin follows $h(t)$ during the other half;
this results in ${\rm OR} \approx 1/2$.
At very high temperatures, on the other hand, 
each spin can take the values $1$ and $-1$
with equal probability, again resulting in the value of {\rm OR} close
to $1/2$, since about half the spins point in the same direction as $h(t)$ 
at a given instant. 
Somewhere in between the spins may follow closely the external driving field, 
giving rise to a peak in the {\rm OR} as a manifestation of SR.
Within the linear response, the OR takes the analytic form~\cite{SRlong,footnote}
\begin{eqnarray} \label{eq:ORanal}
{\rm OR} &=& \frac{1}{2} + \frac{\Omega}{4\pi} \oint dt\, m(t)\,
{\rm sgn}(\cos\Omega t) \nonumber  \\
& = & \frac{1}{2} + \frac{h_0}{\pi}\frac{\tau - 1 }{ 1 + \Omega^2 \tau^2 } , 
\end{eqnarray}
where $\oint$ represents the integration over a full period of the 
driving field in the stationary state and $\tau$ is given by Eq.~(\ref{eq:tau}).
The resulting behavior of the OR with the temperature is shown 
in Fig.~\ref{fig:ORanal}, which reveals most clearly 
the double-peak structure of SR.
Figure~\ref{fig:ORanal}, where the numerical result directly from 
the equation of motion (\ref{eq:glauber}) has also been displayed,
demonstrates that the analytical result in Eq.~(\ref{eq:ORanal})
is rather accurate.
Further shown in Fig.~\ref{fig:ORanal} is that the
analytical result can even be improved by adopting in Eq.~(\ref{eq:ORanal}) 
the (more precise) numerical values of $\tau$ shown in Fig.~\ref{fig:tau} 
instead of those given by Eq.~(\ref{eq:tau}).

From Eq.~(\ref{eq:ORanal}), it is straightforward to obtain the peak position $T_{SR}$ 
of the OR, which is determined by the condition $(d/dT){\rm OR} = 0$. 
This leads directly to the {\em matching condition} between the
external and internal (intrinsic) time scales:
\begin{equation}\label{eq:matching}
\tau = 1 + \sqrt{ 1 + \Omega^{-2} } ,
\end{equation}
which reduces in the limit $\Omega \rightarrow 0$ 
to the direct matching condition between the frequency scales:
\begin{equation}
\tau^{-1} \approx \Omega .
\end{equation}
It is thus concluded that
the resonance becomes strong
when the external time scale $1/\Omega$ 
matches the intrinsic time scale $\tau$ according to the matching 
condition in Eq.~(\ref{eq:matching}).
Since $\tau$, diverging at $T_c$, decreases away from $T_c$,
there in general exist two temperatures, 
one above and the other below the transition,
at which Eq.~(\ref{eq:matching}) is satisfied for given $\Omega$;
obviously, this gives rise to the double peaks.
The inset of Fig.~\ref{fig:ORanal} displays the matching condition
in Eq.~(\ref{eq:matching}), the right-hand side of which is, 
for $\Omega/2\pi =0.1$, depicted by the horizontal dashed line: 
The two intersections determine precisely the
positions of the two SR peaks in the main body of Fig.~\ref{fig:ORanal}, 
demonstrating explicitly the proposed matching condition 
for the existence of the double SR peaks.

\begin{figure}
\twofigures[height=5.8cm]{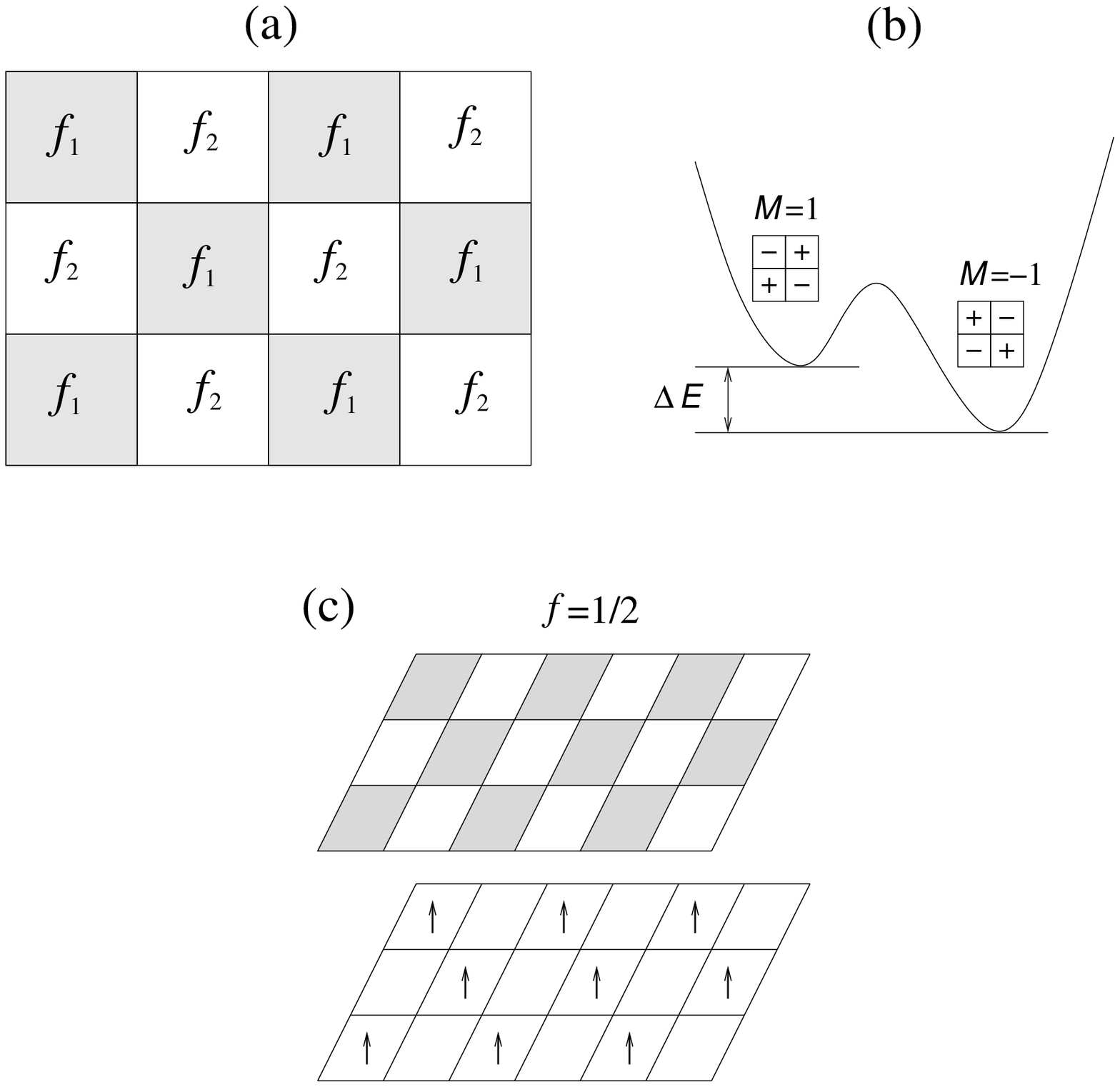}{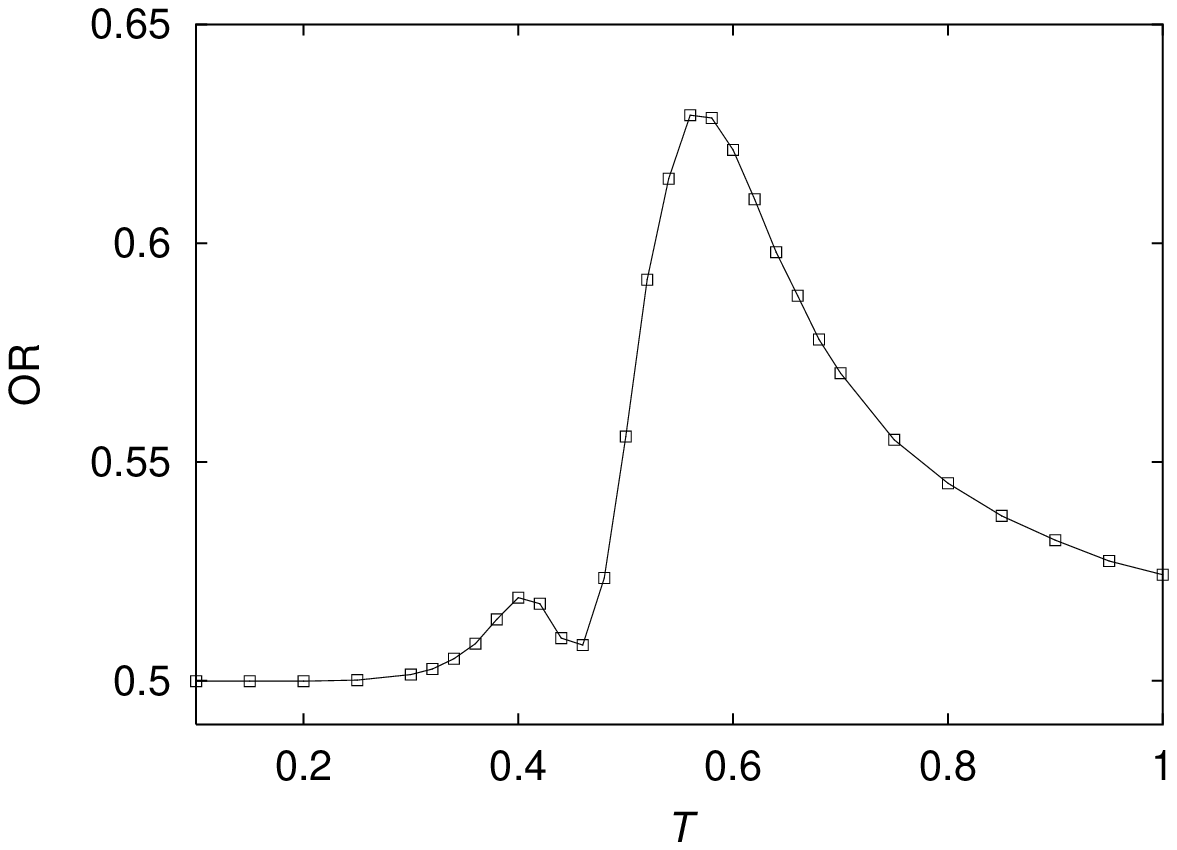}
\caption{
(a) Schematic diagram of the fully-frustrated Josephson-junction
array in the presence of an oscillating staggered
magnetic field: $f_{1,2} = 1/2 \pm f_0 \cos\Omega t$.
Depending on the sign of $\cos\Omega t$, vortices energetically favor
to be located on either shaded plaquettes or empty ones at given
time; the double degeneracy present for $f_0 =0$
is lifted and the system can effectively be regarded as a
two-state system in a double-well energy landscape as
schematically shown in (b), where $M$ denotes the staggered 
vorticity and $\Delta E$ is given in text.
(c) A fully frustrated Josephson-junction array put on the
top of a periodic array of magnetic particles, with either array
driven periodically, provides a possible
experimental realization. 
}
\label{fig:jja}
\caption{
Occupancy ratio versus temperature in a fully frustrated 
two-dimensional Josephson-junction array driven by a weak 
oscillating staggered magnetic field. 
Simulation data for the $16 \times 16$ array 
at $f_0 = 0.02$ and $\Omega=0.01$ are represented by squares, 
whereas the line is merely a guide to the eye.
The double-peak structure is obvious although
the peak below the transition is weaker than the one above.
}
\label{fig:orffxy}
\end{figure}

To disclose ubiquity of the double-peak structure associate with a dynamic phase
transition, we next consider the 2D $L\times L$ FFJJA,
driven by a weak magnetic field, staggered in space and periodic in time,
in addition to the uniform field introducing full frustration
[see Fig.~\ref{fig:jja}(a)]. 
Such a system is described by the Hamiltonian
\begin{equation} \label{XY}
H = - J \sum_{\langle i, j\rangle} \cos (\phi_i -\phi_j -A_{ij}), 
\end{equation}
where $J$ is the Josephson coupling strength, 
$\phi_i $ denotes the phase of the superconducting order parameter 
at site $i$, and the summation is over all nearest neighboring pairs.
The bond angle is given by the line integral of the vector potential
and consists of two parts:
$A_{ij} = A_{ij}^0 + \delta A_{ij}(t)$,
where $A_{ij}^0$ describes the uniform magnetic field
corresponding to half a flux quantum per plaquette ($f=1/2$)
and $\delta A_{ij}(t)$ takes into account 
the additional oscillating magnetic field of frequency $\Omega$.
When the oscillating field is absent ($\delta A_{ij}=0$),
the ground state possesses the two-fold
degeneracy and vortices form a 2$\times 2$ superlattice at zero temperature~\cite{choi}. 
On the other hand, the presence of the oscillating field, 
which leads to the oscillating flux $\delta f(t)= f_0 \cos\Omega t$,
lifts the two-fold degeneracy.
For $\Omega \ll 1$, 
the energy splitting $\Delta E$ between the two states
[each of which has the staggered vorticity $\pm 1$; see Fig.~\ref{fig:jja}(b)]
is given by~\cite{SRlong}:
$ \Delta E  = 2J\sqrt{2} \sin( \pi f_0 \cos \Omega t/2)$.
It is thus clear that this FFJJA subject to the oscillating staggered
magnetic field contains all the basic ingredients
required for SR behavior.
Figure~\ref{fig:jja}(c) suggests one possible realization of this system
in experiment.  Periodic arrays of magnetic particles
are already available and used frequently in experiment~\cite{martin}, 
and we believe that the relevant
SR behavior can be observed through the use of various
vortex imaging techniques.

The dynamics of the system is described by 
the resistively-shunted junction model, stating the current conservation
at each site:
\begin{equation} \label{motion}
{\sum_j}' [ \dot{\phi}_{i}-\dot{\phi}_{j} -\dot{A}_{ij}
  + \sin (\phi_{i}-\phi_{j} -A_{ij} ) + \zeta_{ij}] = 0,
\end{equation}
where the prime restricts the summation to the neighbors of site $i$,
and $\zeta_{ij} (t)$ is the random noise current across the junction with
zero mean and correlation
$\langle \zeta_{ij} (t) \zeta_{kl} (t') \rangle 
= 2 T (\delta_{ik}\delta_{jl} - \delta_{il}\delta_{jk}) \delta(t{-}t').$
Note that energy (and temperature) has been measured in units of $J$ 
and time in units of $\hbar/4e^2 JR$ with the shunt resistance $R$.

To obtain the dynamic behavior, we integrate directly the set of equations of
motion (\ref{motion}) and measure the staggered vorticity at each plaquette. 
The OR of the staggered vorticity, measuring how many plaquettes have
the staggered vorticity preferred by the external driving, is then 
computed according to Eq.~(\ref{eq:OR}),
with the oscillating flux $f(t)$ taking the role of $h(t)$~\cite{SRlong}.
Figure~\ref{fig:orffxy} displays the obtained OR as a function of the 
temperature $T$ in the array of size $L=16$ for $f_0 = 0.02$ and $\Omega=0.01$.
Observed clearly is the emergence of double SR peaks,
one below and the other above the (dynamic) transition temperature 
$T_c \approx 0.443$~\cite{SRlong}, 
although the peak below the transition appears much
weaker than the one above. 
Note that sufficient decrease of $\tau$ is essential for the time-scale
matching condition to be satisfied.  In some systems, 
$\tau$ may become large at low temperatures as well as near $T_c$ and the 
SR peak below $T_c$ may not be observed~\cite{gun}.

In summary, we have proposed the extended time-scale matching condition for
the SR, between the external periodic driving and the relaxation time 
associated with a phase transition,
thus elucidating the link between the appearance
of double SR peaks and the existence of a dynamic phase transition.
This has been demonstrated in the 2D fully frustrated Josephson-junction array
as well as in the mean-field kinetic Ising model.
The argument presented here is rather general, and we
expect that the double-peak structure should appear ubiquitously 
in systems undergoing dynamic phase transitions with diverging
relaxation time at the transition. 

\acknowledgments
This work was supported in part by the Swedish Natural Research Council
through Contract No. FU 04040-332 (BJK and PM), 
by the Ministry of Education of Korea through the BK21 Program (HJK and MYC), 
and by the Korean Science and Engineering Foundation through 
the Center for Strongly Correlated Materials Research (GSJ).


\begin{thebibliography}{0}
\bibitem{footnote}
Our OR is closely related to the measure 
$C = (\Omega/2\pi)$ $\oint m(t) \cos\Omega t$ used
in \cite{leung}.
\end{thebibliography}
\end{document}